# Imaging Antiferromagnetic Domains in Nickel-oxide Thin Films by Optical Birefringence Effect

Jia Xu[1], Chao Zhou[1], Mengwen Jia[1], Dong Shi[1], Changqing Liu[1], Haoran Chen[1], Gong Chen[2], Guanhua Zhang[3], Yu Liang[3], Junqin Li[4], Wei Zhang[5]*, Yizheng Wu[1,6]*

[1] Department of Physics and State Key Laboratory of Surface Physics, Fudan University, Shanghai 200433, China

[2] Department of Physics, University of California, Davis, California 95616, USA

[3] State Key Laboratory of Molecular Reaction Dynamics, Dalian Institute of Chemical Physics, Chinese Academy of Sciences, Dalian, Liaoning, 116023, China

[4] Shanghai Synchrotron Radiation Facility, Zhangjiang Laboratory, Chinese Academy of Sciences, Shanghai, 201204, China

[5] Department of Physics, Oakland University, Rochester, Michigan, USA

[6] Collaborative Innovation Center of Advanced Microstructures, Nanjing 210093, China

*e-mail: weizhang@oakland.edu; wuyizheng@fudan.edu.cn

Abstract:

Recent demonstrations of electrical detection and manipulation of antiferromagnets (AFMs) have opened new opportunities towards robust and ultrafast spintronics devices. However, it is difficult to establish the connection between the spin-transport behavior and the microscopic AFM domain states in thin films due to the lack of the real-time imaging technique under the electric field. Here we report a large magneto-optical birefringence effect with polarization rotation up to 60 mdeg in thin NiO(001) films at room temperature. Such large optical polarization rotation allows us to directly observe AFM domains in thin-film NiO by utilizing a wide-field optical microscope. Complementary XMLD-PEEM measurement further confirms that the optical contrast is related to the NiO AFM domain. We examine the domain pattern evolution at a wide range of temperature and with the application of external magnetic field. Comparing to large-scale-facility techniques such as the X-ray photoemission electron microscopy, using wide-field, tabletop optical imaging method in the reflection geometry enables straightforward access to domain configurations of single-layer AFMs.

1. Introduction

The emerging field of antiferromagnetic (AFM) spintronics is believed one of the most promising contenders for future energy-efficient, high-speed, robust information storage and processing [1-10]. The antiparallel AFM sub-lattices produce zero dipolar fields, making them inert to external magnetic-field perturbations, thus allowing multi-level stability [11,12] in memory devices. In addition, their pronounced exchange anisotropy results in much faster spin dynamics (up to THz) [13-15]. The past decade has witnessed an increasingly active role of AFMs in spintronic device building blocks [1-10], opposing to AFMs' conventional, passive exchange-bias effect [16,17]. However, the absence of a net magnetization in AFMs renders conventional magnetometry ineffective, which cast a great challenge in accessing their magnetic properties especially at the microscopic scale.

To date, the photoemission electron microscopy (PEEM) based on X-ray magnetic linear dichroism effect (XMLD) [2,7,18-21] remains the most commonly adopted technique to study the AFM domains. Such a technique is not widely accessible and is difficult to be incorporated simultaneously with state-of-the-art magneto-transport measurements. Scanning probe techniques such as spin-polarized scanning tunneling microscopy [22], magnetic exchange force microscopy [23] or scanning single-spin magnetometer [24], have been used for AFM imaging but all require well-ordered, pristine sample surfaces. Magneto-optic Kerr effect has also been demonstrated but such mechanism only works for non-collinear AFMs like $Mn_3Sn$ [25]. Only recently, it was demonstrated, utilizing a femtosecond pump–probe method, that the Néel vector dynamics of collinear AFMs, CuMnAs [26] and CoO [27], response to the magneto-optical Voigt effect [26] . Such magneto-optical Voigt effect can be used to develop a general AFM imaging technique simultaneously adaptable to electric and magnetic measurements.

Antiferromagnetic NiO, being one of the most common and natural transition-metal oxides, has recently become a fascinating candidate for investigating many novel spintronic phenomena in AFMs, such as spin Hall magnetoresistance [28-31], spin current transformation [32-35], THz magnons [14,36], and spin-orbit-torque switching [7,8,37]. However, these pioneering results relate closely to theories and postulations for their AFM domain distributions, and therefore, it is greatly needed to experimentally elucidate the behavior of AFM domains under the above scenarios. After the very first demonstration of NiO domain imaging by XMLD-PEEM [18], most subsequent studies on NiO domain imaging have been conducted in NiO bulk samples [20,38,39], and in exchange coupled ferromagnet/NiO bilayers [40-42]. Bulk crystals has also been studied by light transmissions, utilizing the optical birefringence effect induced by the small rhombohedral deformation [43]. Such an optical birefringence effect is magnetic related, since the structure deformation accompanies the AFM ordering of NiO. The magneto-optical birefringence effect can therefore be used to image the AFM domains in thin films.

In this paper, we apply the magneto-optical birefringence effect to image the AFM domains in NiO(001) films grown on MgO at room temperature, using a wide-field, tabletop optical microscope with a reflective geometry, and find that the polarization rotation is as large as 60 mdeg. Our results show that the optical contrasts can be sensitively detected between NiO domains with orthogonal in-plane spin directions. Such optical contrasts increase linearly with the film thickness, and disappear above the Néel temperature ($T_N$). Further, complementary XMLD-PEEM images confirm the AFM origin of the measured optical contrast. The similar NiO AFM domain patterns can be observed even after applying a 9 T field, indicating their robustness against external magnetic field. Our study not only demonstrates a general route towards the facile characterizations of AFM domains by an optical microscope, but also provides new understanding of the spin-canting structures of NiO thin films grown on MgO(001) substrates, which is a crucial component in elucidating many existing issues in spin-transport of NiO-based heterostructures.

2. Sample preparation

The single-crystalline NiO films were grown on MgO(001) substrates in ultrahigh vacuum system by molecular beam epitaxy (MBE) [44,45]. The MgO(001) single-crystal substrates were cleaned with acetone and alcohol, followed by annealing at 600 °C for half an hour inside a ultrahigh vacuum chamber. A 6 nm MgO seed layer was deposited at 500 °C before the NiO growth. The NiO film was then grown by evaporating Ni under an oxygen pressure of $1.0 \times 10^{-6}$ Torr at room temperature. For thickness-dependence measurements, the NiO film was grown into a wedge shape by moving the substrate behind a knife-edge shutter. Finally, the samples were capped with a 5 nm MgO as a protective layer. The film thickness was determined by the deposition rate, which was monitored with a calibrated quartz thickness monitor. Sharp reflection high energy electron diffraction patterns reveal excellent epitaxy growth of NiO film with the lattice relation of NiO[100](001)//MgO[100](001) (Supplemental Materials, Fig. S1).

## 3. Results and discussion

### 3.1. Magneto-optical birefringence imaging of NiO

Magneto-optical birefringence effect is induced by different indices of refraction parallel ($n_{\parallel}$) and perpendicular ($n_{\perp}$) to the magnetization vector in a magnetic material. While most previous measurements on magneto-optical birefringence effect have been performed in the transmission geometry [26,43,46-48], only very few ones were for ferromagnetic films in the reflective geometry [49,50] which shows weak domain contrast. In a normal reflective geometry, the incident, linearly-polarized light has a polarization angle offset $\varphi$ away from the magnetization vector, and the polarization of the reflected light rotates a small polarization angle $\theta_v$, which can be expressed as:

$$\theta_v = \frac{r_{\parallel} - r_{\perp}}{r_{\parallel} + r_{\perp}} sin2\varphi \quad (1)$$

where $r_{\parallel}$ and $r_{\perp}$ are the reflection coefficients for light polarizations along and perpendicular to the magnetization vector, respectively, due to the different indices of refraction $n_{\parallel}$ and $n_{\perp}$. Therefore, the magneto-optical birefringence effect can be used to distinguish two magnetization vectors (or their projected components) that are in the same plane, but orthogonal to each other.

Figure 1(a) shows the geometry of AFM domain imaging utilizing magneto-optical birefringence effect for our study. The AFM domain images were obtained at zero magnetic field using a commercial Evico magneto-optic Kerr microscope equipped with a white-color LED source. Here, we consider two NiO domains with orthogonal in-plane Néel vectors. For a normal-incident, linear-polarized light, if the light polarization is 45° away from the in-plane Néel vector of the NiO spins, the polarization angle $\theta_v$ from the two orthogonal AFM domains will be opposite ($\theta_v$ and $-\theta_v$). Adjusting the analyzer by a small offset angle $\theta$ from the extinction position results in a light intensity (***I***) proportional to $\sin^2(\theta-\theta_v)$ and $\sin^2(\theta+\theta_v)$, respectively, for the two domains. Such an intensity difference gives rise to the optical contrasts that can be detected by the CCD camera, which allows distinguishing the 90° NiO domains. To acquire the magnetic contrast, we measure the two optical images for $I(+\theta)$ and $I(-\theta)$, with the analyzer angle set at $+\theta$ (and $-\theta$), and determine the signal asymmetry, i.e. $I_{\text{asym}} = \frac{I(+\theta)-I(-\theta)}{I(+\theta)+I(-\theta)}$. For the areas with different reflection intensity without polarization rotation, such a treatment will result in the same contrast, thus our treatment can further eliminate the artifacts from surface morphology and single out the magnetic contrasts induced by the polarization rotation. The analyzer angle $\theta$ was chosen to be 7 degrees for most measurements except for the $\theta$-dependent measurements.

Figure 1(b) shows a typical NiO domain image acquired from a 20 nm film with a 40 μm field of view measured at room temperature. Quantitative analysis on the contrast histogram clearly indicates a two-level contrast without any intermediate contributions (Supplemental Materials, Fig. S2). This is because that NiO(001) has the in-plane four-fold symmetry, and there are only two types of the in-plane orthogonal domains detectable with the optical birefringence effect. We further repeated the same measurement with the sample rotated for different in-plane orientations $\varphi$. Figures 1(c) and 1(d) show the images acquired for $\varphi = 135°$ and 90°, respectively. As indicated by Eq. 1, $\theta_v$ should be opposite for $\varphi$ equal to 45° and 135°, and zero for $\varphi = 90°$. Indeed, Fig. 1(c) shows the domain contrast opposite to that in Fig. 1(b), and no significant contrast were observed in Fig. 1(d). Upon a quantitative analysis on the intensity asymmetry, we derive a $\varphi$-dependent contrast of the NiO domains in Fig. 1(e), which exhibits a clear $\sin 2\varphi$ dependence, in good agreement with the expected polarization-dependence of the magneto-optical birefringence effect [26,46-50]. It should be noted that, when $\theta_v \ll \theta$, the image contrast $I_{asym}$ is inversely proportional to the analyzer's offset angle $\theta$, i.e. $I_{asym} \propto \frac{2\theta_v}{\theta}$, as shown in Fig. 1(f). Then, we can derive that the polarization rotation angle $\theta_v$ of such a 20 nm NiO film is as large as 60 mdeg. As a comparison, it is noted that the typical longitudinal Kerr angles from Fe [51] or Co [52] thick films are less than 21 mdeg.

In order to further confirm the antiferromagnetic origin of the observed contrast in Fig. 1, we performed the temperature-dependent measurements using a sample holder with a resistive-heating base. Figures 2(a-d) show the optical contrast for a 20 nm NiO film measured with $\varphi=45°$ at different temperatures. The contrasts decrease with increasing temperature, and vanishes at around $515\ K$, corresponding to the $T_N$ of the film. This value of $T_N$, i.e. $515 \pm 5\ K$, quantified from the image contrast in Fig. 2(i), is close to that for the bulk NiO (523 K). The observed disappearing of contrast at high temperature indicates the origin from AFM ordering of NiO, rather than the surface morphology. Similar temperature-dependent measurement was also conducted for a 9 nm NiO film [Figs. 2(e-h)]. The contrast vanishes at around 501 K [Fig. 2(g)], indicating a $T_N$ of $501 \pm 5\ K$ for 9 nm NiO film. The thickness-dependent $T_N$ is in agreement with the finite-size effect [53]. In addition, the temperature-dependent contrasts of both the 20-nm and 9-nm samples follow a characteristic $<\mathbf{M}>_T^2$ behavior, in good agreement with the principle of magneto-optical birefringence effect [26,46-48], and also with the earlier XMLD spectro-microscopy measurement [18].

We also note that the domain patterns are quite robust upon temperature cycling, including the heating/cooling processes across $T_N$. The observed domain images before and after the thermal annealing appear similar, and only a few percent of domain area can be changed by the thermal annealing (Supplemental Materials, Fig. S6). This observation supplies a direct evidence that the formation of AFM domains is determined by the strongly locked AFM spins due to local strains via the magnetoelastic interaction in epitaxial thin films [18,54,55].

We then investigate the thickness-dependence of the NiO domains, as shown in Fig. 3. The measurement was performed on a NiO-wedge sample with the thickness range of 0-18 nm and the thickness slope of 3 nm/mm. Figures 3(a-c) show the typical domain images with different NiO thicknesses, $d_{\text{NiO}}$, acquired at room temperature with $\varphi = 45°$. As $d_{\text{NiO}}$ decreases, the domain contrast gets weaker and the domain density also becomes smaller. For $d_{\text{NiO}} < 2$ nm, the image contrast is too weak to be convincingly distinguished, indicating that the $T_N$ of ~2 nm NiO film is below room temperature. For $d_{\text{NiO}} > 8$ nm, the NiO is in a stable AFM state, and the domain contrast is nearly linearly dependent on the NiO thickness, see Fig. 3(e). The thickness-dependent experiment was also performed with $\varphi = 135°$ at the same sample area. The measured contrast is opposite to that measured at $\varphi = 45°$.

3.2 Comparative imaging and analysis using XMLD-PEEM

Although the temperature and thickness-dependent measurements already indicate that the observed optical contrast originates from the NiO AFM order, the ultimate way to confirm its AFM origin is to directly compare the optical image with the XMLD-PEEM image. We grew a 10 nm NiO film on MgO(001) substrate, then capped the NiO film with a 1.2 nm Pt film for PEEM measurement. In order to directly compare the optical images and PEEM images, we patterned the sample into a disk shape using photolithography, as shown in Fig. 4(a). The NiO disk is surrounded with a 6 nm thick Pt film, which can minimize the charging effect due to the insulating MgO substrate. The XMLD-PEEM measurement was performed at the XPEEM end-station of beamline 09U in Shanghai synchrotron radiation facility (SSRF). The inset in Fig. 4(a) shows the measured morphology image, indicating the flat surface of the NiO film.

Fig. 4(b) shows the measured $L_2$-edge X-ray adsorption spectrums (XAS) with the in-plane (IP) and out-of-plane (OP) x-ray polarizations, which are normalized by the first $L_2$ peak. In consistent with the previous reports [56,57], the second $L_2$ peak for the OP polarization has the higher intensity than that for the IP polarization, indicating that the effective NiO spin within the x-ray spot aligns perpendicular to the film plane. Figs. 4(e) and (f) show the measured PEEM images for the IP and OP polarizations, respectively. The clear contrast in both PEEM images prove the existence of the AFM domains in NiO/MgO(001) system [27,36]. Usually, the PEEM image with the IP x-ray polarization is sensitive to the in-plane AFM spin components, and Fig. 4(e) with the IP polarization is almost identical to the optical image measured with the magneto-optical birefringence effect in Fig. 4(d). So, this measurement further proves the optical contrast indeed originates from the AFM order of NiO. The contrast in the PEEM images are much smaller than that shown in XMLD spectrum in Fig. 4(b), suggesting a very small in-plane magnetization component of each domain. Interestingly, the PEEM image with the OP polarization in Fig. 4(f) shows the three-level contrast, with most areas containing the similar white-grey patterns as in Fig. 4(e), but there are some regions with the black color, as indicated by the dash circle.

The spin structure of bulk NiO features (111) intra-plane-parallel and inter-plane-antiparallel spin alignments, and the Néel vector is along the $<11\bar{2}>$ directions [18,20,58]. Therefore, there are in total 12 different spin alignments in bulk NiO. For thin-film NiO grown on MgO(001) substrates, the in-plane tensile strain favors AFM spins pointing out-of-plane, as evidenced by XMLD measurements [56,57]. Due to this strain effect, the AFM spins in each domain can be speculated to be along $[112]$, $[\bar{1}12]$, $[1\bar{1}2]$ or $[\bar{1}\bar{1}2]$ directions, with the largest perpendicular components of Néel vector. Thus, the NiO domains in NiO/MgO(001) should be T-domain since the spins in the neighboring domains align in the different (111) surfaces [20,42]. Such four-spin states correspond to the in-plane spin projection directions along $[110]$ or $[1\bar{1}0]$, which gives rise to the opposite magneto-optical birefringence effects, as well as the two-level contrast in the PEEM image with the IP polarization. However, those four-spin states can have three different projection levels along the x-ray polarization direction in the PEEM measurement with the OP polarization, as indicated in Fig. 4(c). So, both PEEM and optical measurements can confirm that the spin structure of NiO on MgO(001) substrate only partially tilts outward from the surface, rather than perpendicular as previously claimed [56,57].

It should be noted that, due to the perpendicular anisotropy induced by the strain effect, the spin canting angle $\theta_S$ between the AFM spins and the [001] axis could be smaller than that in the NiO bulk, and the value of $\theta_S$ is possible to be estimated from the measured PEEM images. It is well known that the XMLD signal is proportional to $\cos^2\theta_P$, with $\theta_P$ being the angle between the AFM spin direction and the x-ray polarization [18,59]. The XAS spectra in Fig. 4(b) shows that the second peak has the contrast of 0.37 between the incident x-rays with the IP and OP polarizations, but even after subtracting the background contribution, the PEEM contrast with the IP polarization in Fig. 4(e) is only ~0.012±0.003. Such small PEEM contrast requires very long integration time to acquire the PEEM image in Fig. 4(e), thus the resolution of PEEM image is greatly suppressed due to the drifting effect during the long-time imaging. Since the XMLD-PEEM contrast with the IP polarization should be proportional to $\sin^2\theta_S$, we can estimate the value of $\theta_S$ to be ~10°±2°. Therefore, our estimation shows that the AFM spin canting angle in NiO/MgO(001) is much smaller than the angle of 35.3° between <112> and [001] in bulk NiO. Our PEEM measurements confirm the measured optical patterns by the birefringence effect are associated with the NiO AFM T-domains. Moreover, the two-level image with weak contrast in the PEEM image with IP polarization in Fig. 4(e) suggests no S-domain existing in the NiO/MgO(001) system, since the S-domains with the Néel vector along <112> axes in all (111) planes can give multi-level image with much stronger contrast. The in-plane direction of NiO AFM spins in each AFM domain can be identified by the XMLD effect with the IP x-ray polarization [60], as indicated by the arrows in Fig. 4(e), therefore, after calibrating with the XMLD effect, the optical contrast due to the birefringence effect can be used to identify the orientation of in-plane component of the Néel vector (Fig. 4(d)).

It has been reported that the AFM spins in NiO thin films could be rotated under strong magnetic field utilizing the spin Hall magnetoresistance effect [28,29]. Here, we further explore the magnetic field effect on the NiO domain structure. We performed domain measurements before and after applying large magnetic field on the sample for more than an hour using a commercial physical property measurement system. In Fig. 5, we show the acquired NiO domain patterns with different thicknesses before and after applying a 9 T magnetic field along the <110> direction. After the field, however, the AFM domain patterns are very similar to those acquired from the as-grown samples. The differential images in Figs. 5(c) and (f) show that the strong field only slightly modifies the domain distributions at certain positions, as the blue areas highlighted by the dash circles. The total areal changed by the applied 9 T field can be estimated, which is only ~1.8% for the 15 nm NiO film and ~0.4% for the 9 nm NiO film, indicating that the AFM domains in such NiO/MgO(001) systems are very robust. The AFM domains changed by the applied field should be attributed to the spin-flop coupling between the field and the NiO AFM spins, thus the in-plane component of NiO Néel vector in the blue areas in Fig. 5 should be perpendicular to the field, i.e. along the [$1\bar{1}0$] direction, in consistent with the orientation determined by PEEM measurement shown in Fig. 4. Interestingly, the T-domains in bulk NiO can disappear for the field larger than 2.5 T [43,58]. Therefore, although the earlier MR measurements suggested that the AFM spins within each domain can be rotated under strong magnetic fields [28,29], most NiO domains would still recover to their stable states after the field is removed. Moreover, it should be noted, that similar NiO domain patterns are observed before and after annealing the sample across the $T_N$ (Supplemental Materials Fig. S6). Our results suggest that the observed domains in thin-film NiO/MgO(001) system should be strongly pinned by the local defects [18], and the domain structures at zero field are difficult to be manipulated, making this particular system, i.e. NiO/MgO(001), likely not the most ideal candidate to be used for switchable memory cells in future AFM spintronics applications. Recent reports shows that the spin-orbit torques can switch the domains in NiO on $SrTiO_3$(001) substrates with relatively small current density [8,37], but for NiO on MgO(001) substrates, higher current density is needed [37]. Such a difference was interpreted by the different heat conductivities in these two systems [37]. On the other hand, different substrate choices could induce different levels of AFM domain wall pinning [61,62], which may possibly influence on the spin-orbit torque switching.

3.3. Origin of the birefringence effect in NiO thin film

Next, we discuss the origin of the observed optical contrast due to the magneto-optical birefringence effect. It is well known that bulk NiO distorts from the cubic structure to a rhombohedral structure below $T_N$, and the contraction axis in an AFM domain should be oriented along one of the <111> crystal axes and perpendicular to the AFM spin axis in each domain [43,57,58]. Such structure distortion strongly couples to the AFM order, and induces the optical birefringence effect, which was used to image bulk NiO domains in the transmission geometry [43]. Our PEEM measurements revealed four possible spin axes in NiO films grown on MgO(001) (Fig. 4(c)), therefore, the structure distortion axis is different in each AFM NiO domain. The different structure distortion in each NiO domains can induce strong birefringence effect, which can contribute to the observed optical contrast in our measurements.

On the other hand, the magneto-optical birefringence effect has been observed in FM films[47,49,50], as well as in the AFM CuMnAs film [26], which are expected to originate purely from the magnetic properties, not from the structure deformations. The NiO S-domains with the spins aligning in the same (111) surface should contain the same lattice distortion, thus the associated birefringence effect in different S-domains is expected to have the pure magnetic origin from the Néel vector. We performed the optical measurements in NiO(111) film grown on MgO(111) surface, but no optical birefringence contrast was observed. Judging from the RHEED patterns, the crystal quality of NiO/MgO(111) is much worse than that of NiO/MgO(001). We noticed that the AFM domain cannot be observed using the optical birefringence effect even for NiO films grown on MgO(001) substrates with poor surface quality, which may indicate very small AFM domain size in the NiO films due to the reduced film quality. As a result, it still requires further investigations both experimentally and theoretically to identify whether the Néel order in NiO thin film can induce significant magneto-optical birefringence effect. Nevertheless, our studies suggest that the magneto-optical birefringence effects from both the magnetostriction effect and the magnetic ordering effect can be applied to image the AFM domains in thin film, which will be relevant to studying AFM materials in the spintronics context. If the sign of the magneto-optical birefringence effect can be calibrated through other measurements, such as XMLD effect or field-dependent effect, the orientation of in-plane component of the Néel vector in each AFM domain can be identified through the contrast of optical imaging.

4. Conclusion

In summary, we demonstrate the magneto-optical detection of antiferromagnetic domains in a collinear AFM thin film, i.e. NiO grown on MgO(001) substrate, utilizing the optical birefringence effect combined with a tabletop, magneto-optical microscopy technique in the reflective geometry. A large polarization rotation up to 60 mdeg at room temperature can be observed for a 20-nm-thick NiO(001) film. By directly imaging the domain patterns, we are able to investigate the evolutions of domain configuration systematically under different thicknesses, magnetic fields, and temperatures. The XMLD-PEEM measurements confirm the AFM origin of the measured optical contrast. We elucidate on the surface spin-canting structure of NiO on MgO(001), and found that the AFM NiO spins are only partially tilted outward from the surface with a tilting angle of ~10°. Finally, our imaging upon repeated temperature and field cycling confirm the strong pinning of the AFM domain walls induced by the film defects. As a tabletop technique, our approach to image the AFM domains utilizing the optical birefringence effect is considerably more accessible than user facility instruments such as XMLD-PEEM, and can also be made adaptable with external magnetic fields or electric currents, which are extremely important in future experiments involving electric- and magnetic-field driven AFM dynamics and switching.

The work at Fudan University was supported by the National Key Basic Research Program of China (Grant No. 2015CB921401), National Key Research and Development Program of China (Grant No. 2016YFA0300703), National Natural Science Foundation of China (Grants No. 11474066, No. 11734006, and No. 11434003), and the Program of Shanghai Academic Research Leader (No. 17XD1400400). Work at Oakland University was supported by the U.S. National Science Foundation under Grants No. DMR-1808892. Work at University of California, Davis was supported by the UC Office of the President Multi-campus Research Programs and Initiatives MRP-17-454963 and NSF DMR-1610060.

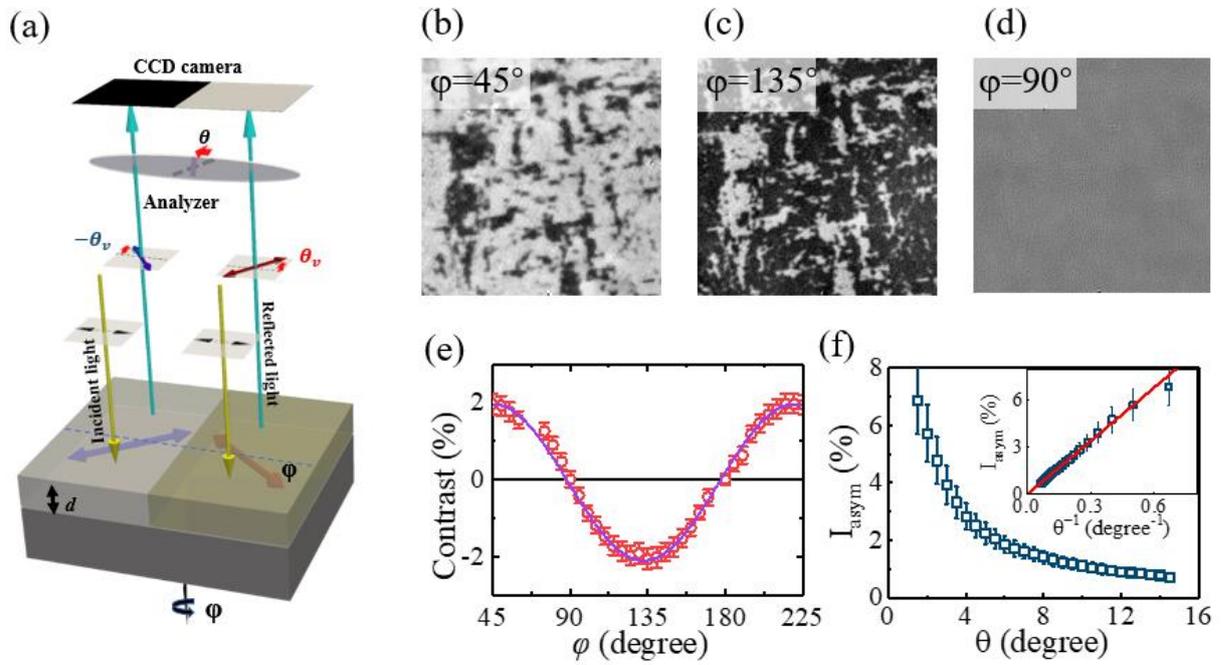

Fig. 1. (a) Schematics of magneto-optical microscopy measurement geometry. (b)-(d) The typical domain images of a 20 nm thick NiO film on MgO(100), obtained from the magneto-optical microscope at room temperature with $\varphi$=45°, 135°, and 90°, respectively. The size for all the images is 40×40 μm². (e) The quantified optical contrasts from the NiO domains as a function of $\varphi$. The purple line represents a fitting curve of sin2$\varphi$. (f) The optical signal symmetry as a function of $\theta$, and the inset demonstrates the linear dependence of $I_{asym}$ as a function of $\theta^{-1}$

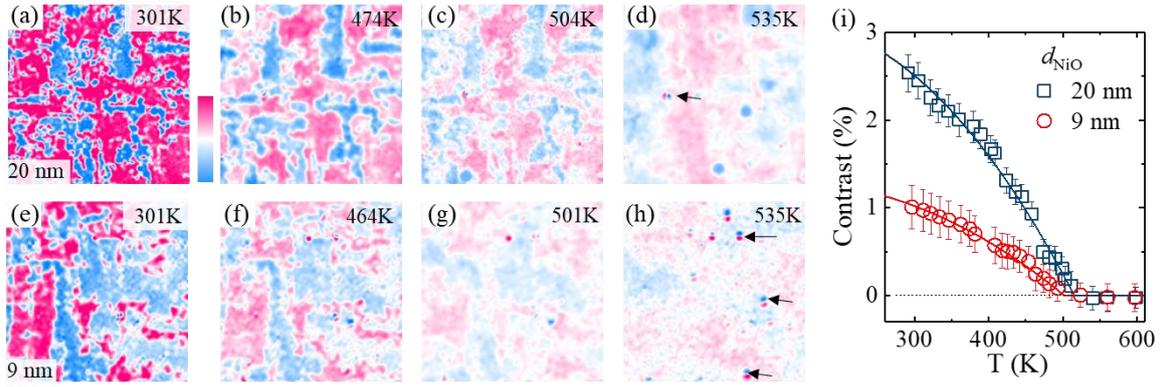

Fig. 2. The temperature dependence of the NiO domains for two NiO thicknesses of (a-d) $d_{NiO}$ = 20 nm and (e-h) $d_{NiO}$ = 9 nm. The weak contrasts in (d) and (h) indicate their non-magnetic origins, and the strong contrasts at the locations indicated by the black arrows in (d) and (h) are due to the surface defects. (i) The optical contrasts as a function of temperature for 9 and 20 nm NiO films, respectively. The solid lines are the theoretical temperature dependence as discussed in the text. The size for all the images is 40×40 μm².

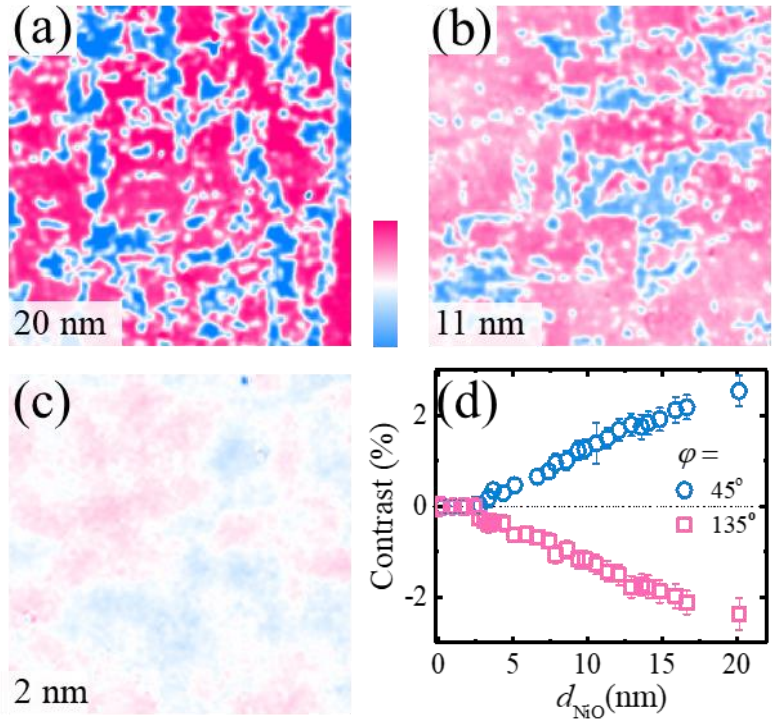

Fig. 3. Thickness dependent measurements of NiO AFM domains. (a-c) The NiO AFM domain images with different film thicknesses measured at $\varphi=45°$. The image sizes are 40×40 µm$^2$. (d) Domain contrast as a function of NiO thickness for $\varphi=45°$ and 135°.

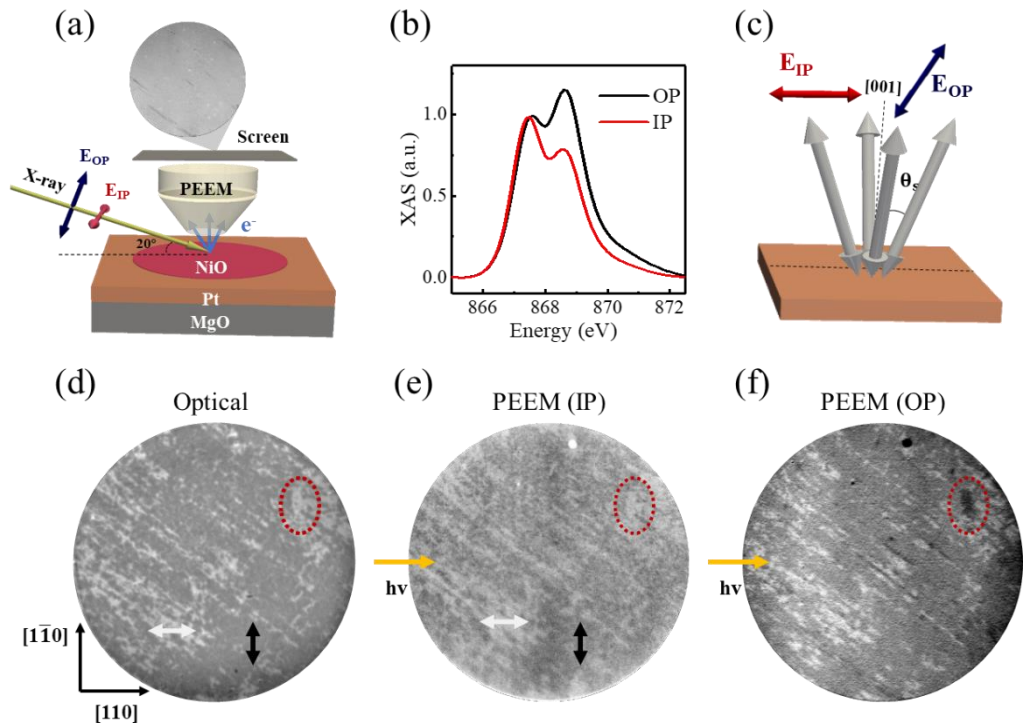

Fig. 4. (a) Schematics of photoemission electron microscopy measurement geometry and the patterned sample of Pt (1.2 nm) / NiO (10 nm) on MgO(001). The inset shows the typical XAS image. (b) XAS spectrums with the in-plane (IP) and out-of-plane (OP) X-ray polarizations from the NiO film. (c) Schematic drawing of four possible NiO spin configurations respect to the OP and IP polarizations. (d) The optical image from the NiO disk with the diameter of 40 μm. (e) and (f) The measured PEEM images from the NiO disk with the OP and IP polarizations, respectively. The arrows in (d) and (e) indicate the orientations of the IP component of NiO Néel vector in the white and black domain areas.

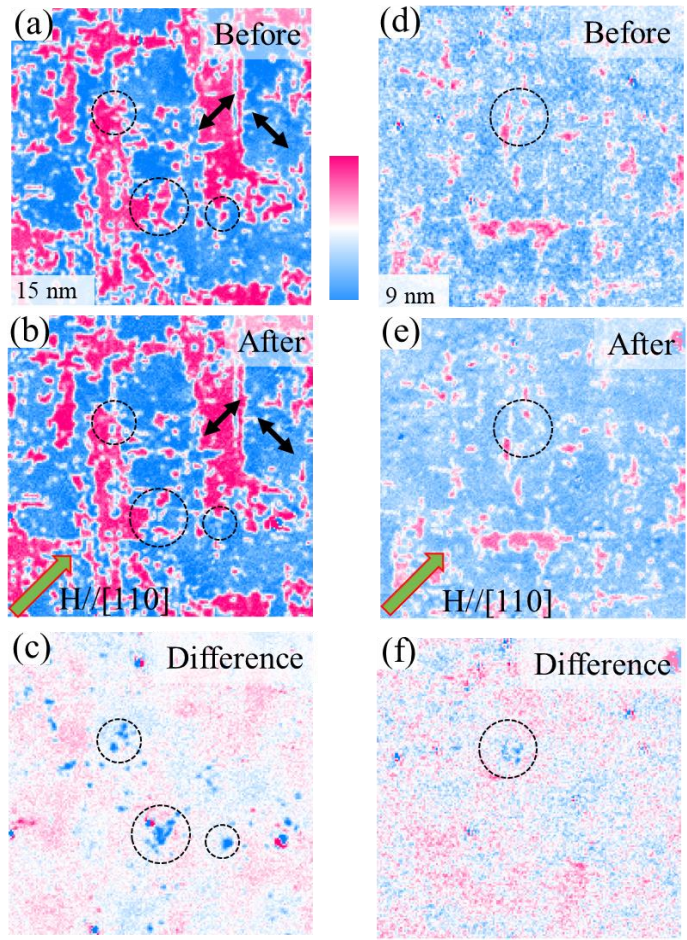

Fig. 5. Effects of the magnetic field on NiO AFM domains. (a-b) The AFM domain images of a 15 nm NiO films (a) before and (b) after applying a 9 T field along the [110] direction. (c) The differential image between (a) and (b), and the dash circles in (a-c) highlight the domain changes after applying the field. The black arrows in (a) and (d) indicate the axis of the IP component of NiO Néel vector in each domain. (d-e) The AFM domain images of a 9 nm NiO film before and after applying a 9 T field along the [110] direction. (f) The differential image between (d) and (e), and the dash circles in (d-f) highlight the domain changes after applying the field. The image sizes are 40×40 μm$^2$.